\documentclass{article}
\def\vp{{\bf p}}
\def\ko{K^0}
\def\kb{\bar{K^0}}
\def\al{\alpha}
\def\ab{\bar{\alpha}}
\def\CPbar{\hbox{{\rm CP}\hskip-1.80em{/}}}
\def\p{\partial}
\def\p{\partial}

\def\g{\gamma}
\def\a_ol{\a_olpha}
\def\d{\delta}
\def\de{\delta}
\def\D{\Delta}
\def\ld{\lambda}

\def\ep{\epsilon}
\def\e{\eta}

\def\om{\omega}

\def\rh{\rho}
\def\s{\sigma}

\def\b{\beta}

\def\a{\alpha}

\def\pdellx'{\frac{\partial}{\partial x'}}
\def\pdellw'{\frac{\partial}{\partial w'}}
\newcommand{\be}{\begin{equation}}
\newcommand{\ee}{\end{equation}}
\def\bed{\begin{displaymath}}
\def\eed{\end{displaymath}}
\def\bea{\begin{eqnarray}}
\def\eea{\end{eqncrray}}
\def\[{$$}
\def\]{$$}
\newcommand{\beas}{\begin{eqnarray*}}
\newcommand{\eeas}{\end{eqnarray*}}

\newcounter{stokes}

\begin{document}
\title{\large Experimental Tests on Yang-Mills Gravity \\
\vspace{0.1in}
with Accurate Measurements of the Deflection of Light}
\bigskip
\bigskip

\author{ Jong-Ping Hsu \\
 Department of Physics,\\
 University of Massachusetts Dartmouth \\
 North Dartmouth, MA 02747-2300, USA}

% Beginning of the text
\maketitle

{\small In the geometric-optics limit, Yang-Mills gravity with space-time translational gauge symmetry predicts $\D \phi =7Gm/(2R) \approx 1.53''$ for the deflection of a light ray by the sun.  The result, which is about 12\% smaller than that in the conventional theory, is consistent with experiments involving optical frequencies that had an accuracy of  $10-20$\%.
%\section{\large Introduction}
%\noindent
\bigskip 

%Keywords: Bending of light experiment, Yang-Mills gravity, effective Riemannian metric tensors.
\bigskip

  %PACS No.: 04.80.Cc,  04.20.Cv}

 \bigskip

\section{Introduction}

In recent papers, we discussed quantum Yang-Mills gravity and a unification of gauge field theories within a generalized Yang-Mills framework with local translational ($T_4$) gauge symmetry in flat space-time.\cite{1,2,3}   The unified model  can accommodate all conservation laws and quantizations of all fields.  The translational gauge symmetry in flat space-time dictates a new gravity-electromagnetic coupling through the gauge covariant derivative.\cite{4}  The unified model is experimentally consistent due to the emergence of  effective Riemann metric tensors in the geometric-optics limit of the electromagnetic and particle wave equations.\cite{5} 

 In the unified model, the electromagnetic gauge curvature (or field strength) involves a very small violation of the $U_1$ gauge invariance due to gravity.
Such a small violation stems from the requirement that all gauge fields and generators associated with the groups $T_4, [SU_3]_{color}, \ SU_2$ and $U_1$ should be on equal footing in the total gauge covariant derivative,
$$
 d_\mu=\p_\mu - ig\phi^\nu_\mu p_\nu+  ieA_\mu+.....,  
 $$
 just like that in the $SU_2\times U_1$ electroweak theory.   In this model, there is a small difference in the effective Riemannian metric tensors between the electromagnetic eikonal equation and the Hamilton-Jacobi type equation derived from wave equations of quarks and leptons in the geometric-optics limit.  The new eikonal equation predicts  a slightly different angle for the deflection of light, which is consistent  with experiments that had an accuracy no better than $10-20$\%.\cite{6}  
 
\section{Effective Riemann metric tensors in Yang-Mills gravity}
 
For the purpose of experimental test, it suffices to concentrate on the $T_4$ gauge invariant action $\int d^4 x L_{em}$ involving only the electromagnetic field $A_\mu$ and a charged fermion $\psi$, which are coupled to the symmetric gravitational field $\phi_{\mu\nu}=\phi_{\nu\mu}$,\cite{1,4} 
 \be
L_{em} = - \frac{1}{4}F_{\mu\nu}F^{\mu\nu} +\frac{i}{2}\left[\frac{}{}\overline{\psi} \g^\mu (\Delta_\mu+ieA_\mu) \psi - [(\Delta_\mu-ieA_\mu)\overline{\psi}] \g^\mu  \psi\frac{}{}\right] - m\overline{\psi} \psi, \ \ \ \ \   
\ee
%%%%1
$$
F_{\mu\nu} = \D_\mu A_\nu - \D_\nu A_\mu, \ \ \  \D_\mu = (\d^\nu_\mu  + g\phi^\nu_\mu) \p_\nu \equiv J_{\mu}^{\nu}\p_{\nu},  \ \ \ \  \phi_{\nu\mu}=\eta_{\ld\nu} \phi^\ld_\mu,
$$
in inertial frames with the metric tensors $\e_{\mu\nu} = (1,-1,-1,-1)$ and $c=\hbar=1$, where $A_{\mu}$ is assumed to satisfy the usual gauge condition $\p^\mu A_{\mu} = 0$.  The $W^{\pm}$ and $Z^o$ gauge bosons and others are not considered  here.  The unified gauge covariant derivative, $d_{\mu}=\D_\mu + ieA_\mu +...=\p_\mu + g\phi^\nu_\mu \p_\nu+  ieA_\mu+...$, involves $p_\nu = i \p_\nu$, which is a representation of the generators of the space-time translational group $T_4$.\cite{2}  

The generalized Maxwell's wave equations can be derived from (1).  We have
\be
 \p_{\a}(J^{\a}_{\mu}F^{\mu\nu}) = e \overline{\psi}\g^{\nu} \psi, \ \ \ \ \ \ \  J_\mu^\a  =\d_\mu^\a + g\phi_\mu^\a.
\ee
%%%%%%2
The wave equation (2) implies a new continuity equation for the electric current in the presence of gravity,
 \be
 \p_{\nu}[e \overline{\psi}\g^{\nu} \psi - g\p_{\a}(\phi^{\a}_{\mu}F^{\mu\nu})] = 0,
 \ee
 %%%%%%%%%%%%3
where we have used the identity $\p_\mu \p_\nu F^{\mu\nu} = 0$.  The first term in (3) is the usual electric current, and the second term is a new `gravity-em current' due to the gravity-electromagnetic coupling in (1).  It involves a factor $g\phi^{00}\approx g\phi^{11}\approx Gm/r$, which is roughly $10^{-9}$ on the surface of the earth.  Such a small effect is difficult to detect on the earth.
 
To test Yang-Mills gravity, let us concentrate on the effective metric tensors in the geometric-optics limit of wave equations for quantum particles.  Yang-Mills gravity is formulated within the framework of flat space-time.  However, in the geometric-optics limit, the $T_4$ gauge invariant Dirac wave equation for a charged fermion (e.g., quark and lepton) derived from (1) lead to the following Hamilton-Jacobi type equation with the same effective metric tensor $G^{\mu\nu}$,\cite{1} 
\be
 G^{\mu\nu}(\p_{\mu}S)(\p_{\nu}S) - m^{2} = 0,  \ \ \ \ \ \ \  G^{\mu\nu} = \e_{\rh\ld} J^{\rh\mu} J^{\ld\nu},
\ee
%%%%%%%%%%%%%%%%%%%5%%%4
where we have used the limiting expression for the fermion field $\psi=\psi_o exp(iS)$ and the momentum $\p_{\mu} S$ and mass $m$ are very large.\cite{7,1} We stress that the equation of  motion for classical objects in Yang-Mills gravity is derived from the $T_4$ gauge invariant wave equation for quantum particles (in the presence of gravitational field) by taking the geometric-optics limit, in contrast to that in the conventional theory of gravity. 

The fundamental equation for a light ray in geometric optics can also be derived from the translational gauge invariant action.  For electromagnetic waves, we have the usual limiting expression for field,\cite{7} $A_\mu=a_\mu exp({i\Psi})$,  where the wave 4-vectors $\p_\mu \Psi$ are very large.  We also assume that $a_\mu$ can be expressed in terms of the space-like polarization vector $\ep_{\mu}(\ld)$, i.e., $a_\mu = \ep_\mu(\ld)b(x), \ b(x) \ne 0$.  As usual, we have  $\sum_{\ld} \ep_{\mu}(\ld)\ep_{\nu}(\ld) \to -\e_{\mu\nu}$ by taking the polarization sums.\cite{8}
 Since we are interested in the propagation of light in vacuum, let us consider the generalized Maxwell's wave equation (2) with the gauge condition  $\p_\mu A^\mu = 0$ and $e=0$.  In the geometric-optics limit, equation (2) leads to
\be
Z^\rh_\mu a^\mu = 0,  \ \ \ \ \ \ \ \ \   
Z^\rh_\mu = G^{\s\ld} \p_\s \Psi \p_\ld \Psi \de^\rh_\mu  - g\phi^\ld_{\mu}J^{\rh\s} \p_\s \Psi \p_\ld \Psi.
\ee
%%%%%%%%%%%%5
The first term in $Z^\rh_\mu$ is the larger one and leads to  the same as the effective metric tensor $G^{\mu\nu}$ for the motion of classical objects in (4).  The second term in $Z^\rh_\mu$ was smaller than the first term and was not considered in previous discussions.\cite{1,2}   

It turns out to be interesting to investigate these two terms because the smaller second term in (5) suggests an observable departure from the conventional theory for the deflection of light.  Multiplying $Z^\rh_\mu a^\mu $ in (5) by $a^\nu \e_{\nu\rh}$ and taking the polarization sums, we obtain 
\be
\frac{1}{b^2 (x)}\sum_{\ld}Z^\rh_\mu a^\mu a^\nu \e_{\nu\rh }= -  \de_\rh^\mu Z^\rh_\mu=0.
\ee
%%%%%%6
This equation, together with $Z^\rh_\mu $ in (5), enables us to calculate the contribution of the non-dominate term to the deflection of light.  After some calculations, we find a new eikonal equation, which involves a slightly different effective metric tensor $G_L^{\mu\nu}$ for the motion of a light ray,
\be
G_L^{\mu\nu} \p_\mu \Psi \p_\nu \Psi  = 0,
\ee
%%%%%%%%3%%%6%%%%%%7
$$
G_L^{\mu\nu}  = G^{\mu\nu} -\frac{g}{4}\phi_{\ld}^\mu J^{\ld\nu}
=\e_{\a\b}(\e^{\mu\a} + g \phi^{\mu\a})(\e^{\nu\b} + \frac{3g}{4} \phi^{\nu\b}),
$$
%%%%%%%%%%%%%%4%%%%%%%7%%%%%
in the geometric-optics (or classical) limit.   We note that the effective Riemannian metric tensor $G_L^{\mu\nu}$ for light rays also appears in the classical limit of  wave equations for other vector gauge bosons (i.e.,  the $Z^o$ boson and the $W^{\pm}$ bosons) in the unified model.\cite{5}  

\section{Experimental tests of Yang-Mills gravity}

For experimental test, let us consider the new eikonal equation (7) and the static gravitational potential.  In the spherical coordinates,  the static fields produced by the sun with mass $m$ in Yang-Mills gravity are\cite{4}\footnote{We have used symbolic computing (xCoba in xAct by D. Yllanes and J. M. Martin-Garcia) to check the approximation solution (8) to order $(Gm/r)^7$ based on the $T_4$ gauge invariant action and the gauge-fixing terms in ref. 4. I would like to thank D. W. Yang and J. Westgate for their help.}
\be
g \phi^{00}= \frac{G m}{r} + \frac{G^2 m^2}{2r^2}, \ \ \ \ 
g \phi^{11}= \frac{G m}{r} + \frac{G^2m^2}{6 r^2},   
\ee
%%%%%%%%%%%6%%%%%8
$$
g \phi^{22} =  \frac{1}{r^2} \left[\frac{G m}{r} + \frac{4 G^2m^2}{ 
3 r^2}\right],  \ \ \  g \phi^{33} = \frac{g \phi^{22}}{sin^2 \theta},
$$
where $G=g^2/(8\pi)$ is the Newtonian constant.

  For the perihelion shift for one revolution, equations (4) and (8) lead to the same result as the conventional theory, $\d \phi  = 6\pi Gm/P, \ P=M^2/(m^2_{p} G m)$,\cite{4,1,7} where $M$ is the constant angular momentum of the planet and $m$ and $m_p$ are respectively the mass of the sun and  the planet.  The higher order correction terms to $\d \phi= 6\pi Gm/P$   in Yang-Mills gravity are too small to be detected.\cite{1} 
  
Following   the usual procedure,\cite{7,2} new eikonal equation (7) and the static solutions (8) lead to 
\be
\D \phi_{light} = \frac{7Gm}{ 2R_o}\approx 1.53'',
\ee
%%%%%%%%%%%%%%%%%7%%%9
where $R_o$ is the distance from the center of the sun.  There are higher order corrections  to $\D \phi_{light}$  in (9), but they are too small to detect.\cite{1} 

The result (9) is smaller  than the usual value  $1.75''$ in the conventional theory of gravity by about 12\%.\cite{9,7}  The comparison of theoretical results between the conventional theory of gravity and the Yang-Mills gravity should be made with caution.  The reason is that the calculations in Yang-Mills gravity are carried out in inertial frames, while the corresponding result in the conventional theory is not calculated in the inertial frame.\cite{10}

In contrast to (9), the red shift in Yang-Mills gravity based on flat space-time is the same as the conventional result.  The reason is as follows:  One can say that a photon in a gravitational field has the kinetic energy $\hbar\om$ and potential energy $\hbar\om g\phi^{00}$.\cite{9}  In Yang-Mills gravity, the conservation of energy,\cite{11}\footnote{ In her famous 1918 paper, Noether also has `Theorem II,' which shows that general relativity does not have the conservation law of energy because the group of general coordinate transformations in curved space-time has a continuously infinite number of generators, in contrast to the Lorentz and the Poincar\'e groups in flat space-time.  Yang-Mills gravity based on flat space-time has  the conservation law of energy.}  $\hbar\om + \hbar\om g\phi^{00}=$constant, and the static potential (8) lead to the usual gravitational red-shift, $\om_2/\om_1= (1+g\phi^{00}_1)/(1+ g\phi^{00}_2)$, which has been confirmed to 1\% accuracy.\cite{6,9}  Moreover, Yang-Mills gravity leads to quadrupole radiations, which are consistent with experiments.\cite{12}

\section{Discussions}

The effective Riemann metric tensors in (4) and (7) are derived from the wave equations of fermions and gauge bosons, and they emerge in and only in the limit of the geometric-optics.  Therefore, these classical equations of motion in Yang-Mills gravity hold only  for macroscopic objects and light rays with sufficiently high frequencies.  They may not be applicable to experiments involving, say, radio frequencies.  The difference between $G_L^{\mu\nu}$ for light rays (and massive `gauge-boson rays') and $G^{\mu\nu}$ for quarks and leptons are due to their different  $T_4$ gauge invariant couplings to $\phi_{\mu\nu}$ and their different physical properties of fields.  These are important differences between Yang-Mills gravity (based on flat space-time) and the conventional theory of gravity. 

The reason for the smaller effect of light deflection in (9) is due to the presence of the second term $- (g/4)\phi_\ld^\mu J^{\ld\nu}$ in the effective metric tensor $G_L^{\mu\nu}$ in (7).  Let us expressed this departure of light deflection from Einstein gravity in terms of the post Newtonian parameters $\a,\b,$ and $\g$.  In the geometric-optics limit, the effective metric in Yang-Mills gravity can be defined as $ds^2=I^L_{\mu\nu} dx^\mu dx^\nu$, where $I^L_{\mu\ld} G^{\ld\nu}_L = \de^\nu_\mu$.  Using $G^{\mu\nu}_L$ in equation (7), the static solutions in (8) and the spherical coordinates $(w,r,\theta, \phi)$, Yang-Mills gravity predicts the effective metric in the usual isotropic form to be
\be
ds^2=\left(1-2\a\frac{Gm}{r} + 2\b \frac{G^2m^2}{r^2}+...\right)dw^2
\ee
%%10
$$
 - \left(1+2\g\frac{Gm}{r}+...\right)\left[dr^2 + r^2 d\theta^2 + r^2 sin^2\theta d\phi^2\right],
$$
where
\be
\a=\frac{7}{8}, \ \ \  \b=\frac{23}{32}, \ \ \ \   \g=\frac{7}{8},
\ee
%11
for the motion of a light ray with optical frequencies.  One the other hand, for the motion of macroscopic objects, we use (4), (8) and the definition for $ds^2$, to obtain the result
\be
\a=\b=\g=1,
\ee
%%%%%%%%%%%%12
which is consistent with that in the conventional theory of gravity. 

The post-Newtonian parameters in (11) and (12) do not depend on optical frequencies of light rays.  The reasons are as follows:  Before one takes the geometric-optics limit, the generalized Maxwell's equations (2) in the presence of gravity are very complicated when the usual expression $A_\mu =a_\mu \ exp(i\Psi)$ is used.  The differential equation of the eikonal $\Psi$ does not have the simple Hamilton-Jacobi form and one does not have effective Riemannian metric tensors.  Only when one takes the geometric-optics limit, does one obtain the Hamilton-Jacobi type equation for the wave vector $\p_\mu \Psi$, as shown in the new eikonal equation (7).  In equation (7),  the effective metric tensors $G_L^{\mu\nu}$ turn out to be functions of the gravitational fields $\phi_{\mu\nu}$ only, and do not involve the wave vector $\p_\mu \Psi$ or the frequency of the electromagnetic waves.  Such a simplicity of the eikonal equation appears to be related to the translational gauge symmetry of the action involving the Lagrangian (1).  Therefore, when one uses the effective metric tensor $G_L^{\mu\nu}$ to write down the effective interval $ds^2$ in (10), the post-Newtonian parameters do not depend on the electromagnetic wave vector and, hence, do not depend on the frequencies of light rays.  

 So far, the experimental accuracy for the deflection of visible light is inadequate to test the prediction (9).\cite{9}  One needs a better experimental accuracy to test Yang-Mills gravity.   
 
 There are data of light deflection related to quasars.  They appear to be not accurate enough to test Yang-Mills gravity in the geometric-optics limit.  The vast amount of data are related to the deflection of light by gravitational lenses.  The angle of deflection depends on the mass of galaxies between the quasars and the Earth.  One usually use the result of deflection to estimate the total mass of the foreground galaxy.  However, there is one `relevant' experiment which measured deflection of light from the quasar (J0842+1835) by the Jupiter by using radio-telescope at 84GHz.\cite{13}  The result is (0.98 $\pm$ 0.19)$\times$ (value predicted by GR).  This experiment involves radio frequency $84GHz$, which is not high enough to test result obtained for optical frequencies $\approx 500THz$ in Yang-Mills gravity.  The uncertainty is about $20\%$.  There are more accurate experiments, but they use radio frequencies rather than optical frequencies.  For example, the Shapiro experiment of radar echo delay used the radar frequency of 7840MHz, which is much smaller than the optical frequencies.  So the results of the Shapiro experiments and the data related to quasars are not suitable to test Yang-Mills gravity in the geometric-optics limit.
  
 The significances of such experiments are (i) to reveal whether the Yang-Mills `gauge curvature' of the local translational gauge symmetry or the conventional `space-time curvature'  dictates the gravitational interaction, and (ii) to indicate whether unification of different interactions of quantum fields can be realized in the generalized Yang-Mills framework based on flat space-time.\cite{4,3}  Furthermore,  as stressed by Dyson, there is incompatibility ``between EinsteinÕs principle of general coordinate invariance and all the modern schemes for a quantum-mechanical description of nature.''\cite{14,1}  Accurate measurement of deflection of light will also indicate whether the Yang-Mills gravity in flat space-time is an alternative solution to Dyson's suggestion of some quantum-mechanical analog of Riemannian geometry to resolve 
``the most glaring incompatibility of concepts in contemporary physics.'' 

The work is supported in part by JingShin Research Fund of the UMassD Foundation.  The author would like to thank Leonardo Hsu for earlier collaborations.

%\section*{References}

\bibliographystyle{unsrt}

\end{document}